\documentclass[aps,a4paper,twoside,twocolumn,secnumarabic,balancelastpage,amsmath,amssymb,superscriptaddress 
]{revtex4}

\usepackage{graphics}      
\usepackage[pdftex]{graphicx}      
\usepackage{longtable}     
\usepackage{bm}            
\usepackage{verbatim}
\usepackage{mathtools}
\usepackage{adjustbox}
\usepackage[colorlinks=true]{hyperref}

\usepackage{dsfont}

\usepackage{stmaryrd}

\begin{document}

\title{Non-unique connection between bulk topological invariants and surface physics}

\date{\today}

\author{Corentin Morice}
\affiliation{Center for Electronic Correlations and Magnetism, Theoretical Physics III, Institute of Physics, University of Augsburg, 86135 Augsburg, Germany}
\author{Thilo Kopp}
\affiliation{Center for Electronic Correlations and Magnetism, Experimental Physics VI, Institute of Physics, University of Augsburg, 86135 Augsburg, Germany}
\author{Arno P. Kampf}
\affiliation{Center for Electronic Correlations and Magnetism, Theoretical Physics III, Institute of Physics, University of Augsburg, 86135 Augsburg, Germany}

\begin{abstract}
At the heart of the study of topological insulators lies a fundamental dichotomy: topological invariants are defined in infinite systems, but surface states as their main footprint only exist in finite systems. In the slab geometry, namely infinite in two planar directions and finite in the perpendicular direction, the 2D topological invariant was shown to display three different types of behaviour. The perpendicular Dirac velocity turns out to be a critical control parameter discerning between different qualitative situations. When it is zero, the three types of behaviour extrapolate to the three 3D topologically distinct phases: trivial, weak and strong topological insulators. We show analytically that the boundaries between types of behaviour are topological phase transitions of particular significance since they allow to predict the 3D topological invariants from finite-thickness transitions. When the perpendicular Dirac velocity is not zero, we identify a new phase with surface states but no band inversion at any finite thickness, disentangling these two concepts which are closely linked in 3D. We also show that at zero perpendicular Dirac velocity the system is gapless in the 3D bulk and therefore not a topological insulating state, even though the slab geometry extrapolates to the 3D topological phases. Finally, in a parameter regime with strong dispersion perpendicular to the surface of the slab, we encounter the unusual case that the slab physics displays non-trivial phases with surface states but nevertheless extrapolates to a 3D trivial state.
\end{abstract}

\maketitle

\section{Introduction}

Topological insulators \cite{Hasan2010, Qi2011, Fu2007a, Moore2007} have left the realm of theoretical curiosities to enter the stage of ubiquitous materials; indeed, a significant proportion of known materials has recently been catalogued as topologically non-trivial \cite{Vergniory2019}. The formalisms used to study topological states thereby face their applicability to real-world materials. Topological invariants are defined in reciprocal space, but their imprint has to be necessarily sought for in finite samples. Is the absence or presence of surface states the only trace of the topology of the system? The latter question can be explored, for example, in a slab geometry, which is infinite in two dimensions and finite in the third. A certain number of 2D layers can be stacked to obtain a specific thickness, and taking the limit towards a large number of layers allows to extrapolate to the 3D infinite system size limit. Via this strategy we can explore whether the stacking of topologically trivial layers extrapolates to a 3D trivial state or whether the stacking of quantum spin Hall insulators extrapolates to a 3D topological insulator.

Related is the issue of the bulk-boundary correspondence \cite{BernevigBook}, which states: if two half-spaces with different topological indices are put in contact, there will be gapless states at the boundary. Though, strictly speaking, the topological indices referred to are not the ones of the half-spaces, but of two infinitely extended systems corresponding to each half-space. One could ask: is it possible to reformulate this statement using properties intrinsic to the two half-spaces? Here, we study intrinsic properties of a slab in order to find to what extent they give us information about the topology of the system.

Following the discovery of a strong topological insulating state in Bi$_2$Se$_3$ \cite{Xia2009} and the setting up of a material-specific two-orbital model \cite{Zhang2009, Liu2010a}, the thickness dependence of the spectrum in a slab geometry was calculated \cite{Linder2009, Lu2010, Hao2011}. The gap of the surface states in these spectra decays exponentially but oscillates with thickness. The 2D topological number associated to a slab was also found to oscillate with thickness. The same wavelength is also detected in the spatial decay of the wave-functions of the surface states \cite{Liu2010}. This was also confirmed in ab-initio calculations.

Depending on the choice of parameters, the exponential decay of the surface gap occurs either with or without an oscillatory component \cite{Ebihara2012}. In the former case, the periodicities of the oscillations of the gap and of the 2D topological invariant coincide, while in the latter case the 2D topological invariant simply alternates with every added layer.

The effective model as originally designed for Bi$_2$Se$_3$ was later generalised to the Wilson-Dirac model, whose topological phase diagram in the 2D and 3D limits was calculated \cite{Imura2012, Yoshimura2014}. In a 1D version of the Wilson-Dirac model it was confirmed analytically that the surface state energy gap determines the spatial profile of the surface wave-function \cite{Okamoto2014}, and that two distinct regimes of exponential decay of the surface gap exist.

The evolution of the 2D topological invariant $\nu$ with thickness in a slab geometry was shown to feature three different regions as a function of the Dirac mass \cite{Kobayashi2015, Yoshimura2016}: an ordinary insulating (OI) region where $\nu=0$ at any thickness, a `stripe' region where $\nu=1$ and $\nu=0$ for odd and even number of layers respectively, and between these two a `mosaic' region where $\nu$ oscillates with thickness. The two latter where related to the two regimes of exponential decay of the surface gap: without and with oscillations. For a fixed number of layers, the topological phase transitions as a function of the model parameters were calculated analytically. Finally, the `stripe' and `oscillating' regions where found to extrapolate to 3D weak and strong topological insulating phases respectively, when the perpendicular Dirac velocity was set to zero. Away from zero, the 2D and 3D topological phase transitions were found to be hard to link.

Recent progress in the synthesis of thin films of topological materials allows to establish the connection to experiments. The decrease of the gap in the electronic spectrum upon increasing the thickness of a Bi$_2$Se$_3$ film was measured using angle-resolved photoemission spectroscopy \cite{Zhang2010}. In (Bi$_{1-x}$In$_x$)$_2$Se$_3$, the doping-driven topological phase transition changes with thickness, in accordance with numerical calculations for slab geometries \cite{Salehi2016}. The layered compound ZrTe$_5$ also sparked a lot of interest as a material predicted to be close to a topological phase transition between strong and weak topological insulating states \cite{Weng2014, Fan2017}. A thickness-driven transition between a topological semimetallic state and a conventional metal was conjectured from resistivity measurements \cite{Lu2017}. Moreover, the strong in-plane Hall resistivity in this material, considered to be a signature of a semimetal, vanishes in very thin films \cite{Li2018}. Thickness appears to be a key parameter in the study of topological states.

\begin{figure}
\centering
\begin{tabular}{cr}
a) & \raisebox{-0.95\height}{\includegraphics[width=8cm]{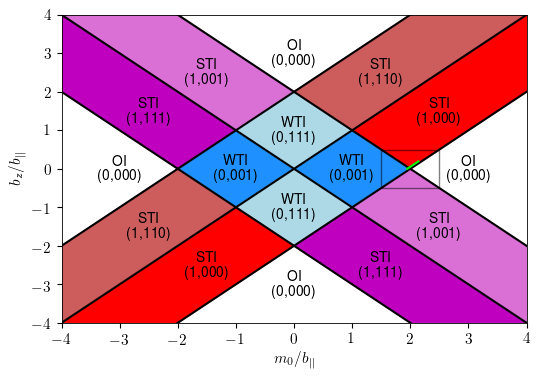}}
\\
b) & \raisebox{-0.95\height}{\includegraphics[width=7.54cm]{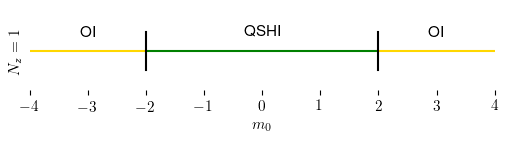}}
\end{tabular}
\caption{Topological phase diagram of the 3D Wilson-Dirac model as a function of $m_0/b_\parallel$ and $b_z/b_\parallel$ (a), and topological phase diagram of the 2D Wilson-Dirac model as a function of $m_0/b_\parallel$ (b), introduced in Ref.\ \cite{Yoshimura2016}. OI, STI and WTI stand for ordinary (trivial) insulator, strong topological insulator, and weak topological insulator, respectively. The numbers in parentheses are the 3D topological indices $\left( \nu_0, \nu_1 \nu_2 \nu_3 \right)$. The green line in the 3D phase diagram is the projection of the phase diagram of the minimal model for Bi$_2$Se$_3$ for the parameter range studied in Ref.\ \cite{Ebihara2012}.}
\label{fig:Bi2Se3-Wilson-3D-phase-diagram-comparison}
\end{figure}

Here, we aim to understand better the crossover between 2D and 3D topological insulating states, by precisely characterising the regions in the thickness-controlled topological phase diagram for a slab geometry. We introduce the model in Sec.\ \ref{Wilson-Dirac model}, and calculate the topological index $\nu$ as a function of thickness in Sec.\ \ref{Z2-invariant maps}, and thereby identify two previously unexplored phases. In Sec.\ \ref{Decomposition of Z2-invariant in parity eigenvalues}, we decompose $\nu$ into parity eigenvalues, which allows to establish a relation between parity cubes in 3D and parity squares in 2D. We show analytically that the outermost transitions of `mosaic' regions extrapolate to topological phase transitions at TRIM points in 3D. In Sec.\ \ref{Impact of tz}, we study the influence of the perpendicular Dirac velocity on $\nu$ and the surface gap, which changes from an oscillating to a non-oscillating exponential decay as a function of thickness for large $t_z$. In Sec.\ \ref{Band inversion}, we explain that the topological phase transition in Sec.\ \ref{Decomposition of Z2-invariant in parity eigenvalues} switches between band inverted and non-band inverted spectra, and therefore that there is a parameter regime without band inversion at any thickness that converges towards a strong topological insulating phase. In Sec.\ \ref{Influence of tz in the 3D limit}, we study how the perpendicular Dirac velocity controls the band structure of the 3D system. Finally, in Sec.\ \ref{Ordinary insulating phase at high bz}, we explore in details a new region in the topological phase for the slab, and explain its origin using the picture defined in Sec.\ \ref{Decomposition of Z2-invariant in parity eigenvalues}.

\section{Wilson-Dirac model}
\label{Wilson-Dirac model}

\subsection{3D Hamiltonian}

We begin by introducing the three-dimensional Wilson-Dirac model, which is a generalisation of the lattice minimal model for Bi$_2$Se$_3$ \cite{Zhang2009, Liu2010a, Yoshimura2016}:
\begin{align}
H(\textbf{k}) = &\left( m_0 - \sum_{\mu = x, y, z} b_\mu \cos(k_\mu) \right)\beta \nonumber
\\
&+ \sum_{\mu = x, y, z} t_\mu \sin(k_\mu) \alpha_\mu
\label{eq:3D-Wilson-model}
\end{align}
where we use the following Dirac matrices:
\begin{equation}
\alpha_\mu = \tau_x \otimes \sigma_\mu, \quad \beta = \tau_z \otimes \mathds{1}_2
\end{equation}
where $\tau_\mu$ and $\sigma_\mu$ are Pauli matrices in the orbital and spin spaces, respectively.
We consider in-plane isotropy and set: $b_\parallel \equiv b_x = b_y$.

This Hamiltonian being inversion-symmetric, its topological phase diagram is obtained by calculating parity eigenvalues of the occupied bands at the time-reversal invariant momenta (TRIM) \cite{Fu2007b}. At these TRIM points $k_x, k_y, k_z \in \{ 0, \pi \}$, the second term in Eq.\ \eqref{eq:3D-Wilson-model} is zero, and the Hamiltonian is diagonal. The topological invariants of the system are therefore entirely determined by the two ratios $m_0/b_\parallel$ and $b_z/b_\parallel$. Since $t_x/b_\parallel$ and $t_y/b_\parallel$ do not influence the topology of the system, we set them to 1 in the following. Neither does $t_z/b_\parallel$, which we denote the perpendicular Dirac velocity, but it will play a role in the case of the slab geometry, and we will specify its value when necessary.

The topological phase diagram of the 3D Wilson-Dirac Hamiltonian was constructed in Ref.\ \cite{Yoshimura2016}. We reproduce this phase diagram in Fig.\ \ref{fig:Bi2Se3-Wilson-3D-phase-diagram-comparison}a, and ---for the purpose of comparison--- superpose the parameter range studied in Ref.\ \cite{Ebihara2012}, considered relevant for Bi$_2$Se$_3$.

The two-dimensional Wilson-Dirac model is the same as the three-dimensional one but the sums run only on $x$ and $y$. Therefore there are no $b_z$ nor $t_z$ parameter in this case. The topological phase diagram of the 2D Wilson-Dirac model (Fig.\ \ref{fig:Bi2Se3-Wilson-3D-phase-diagram-comparison}b) is much simpler than the 3D one, since it only depends on $m_0/b_\parallel$: for $-2<m_0/b_\parallel<2$, the system is a quantum spin Hall insulator, and for $|m_0/b_\parallel|>2$ it is a trivial insulator.

Using the 2D and 3D phase diagrams, we can already address the question: will a stack of topologically trivial layers extrapolate to a 3D trivial state? It can extrapolate towards both a trivial or a strong topological 3D phase. For example, at $m_0/b_\parallel=3$, the 2D system is trivial and the 3D system can be either trivial (e.g.\ at $b_z/b_\parallel=0$), or a strong topological insulator (e.g.\ at $b_z/b_\parallel=2$). Similarly, when the system is topological in 2D, for example at $m_0/b_\parallel=1$, it can be trivial, or weak or strong topological in 3D. The only case which is not possible given these topological phase diagrams is to have a trivial state in 2D and a weak topological insulator in 3D.

Now that we have seen that there does not seem to be a general rule of what one finds in 3D given what one started with in 2D, we define the Hamiltonian in the slab geometry to explore how the system evolves from one to the other.

\subsection{Slab model}

\begin{figure*}
\centering
\begin{tabular}{cccc}
\rotatebox{90}{$b_z=0.01$} & \raisebox{-.5\height}{\includegraphics[width=8cm]{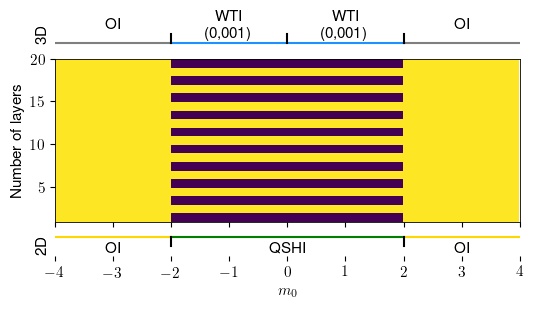}}
& \raisebox{-.5\height}{\includegraphics[width=8cm]{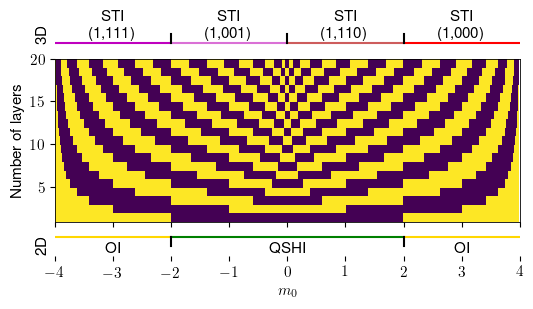}} & \rotatebox{90}{$b_z=2$}
\\
\rotatebox{90}{$b_z=0.5$} & \raisebox{-.5\height}{\includegraphics[width=8cm]{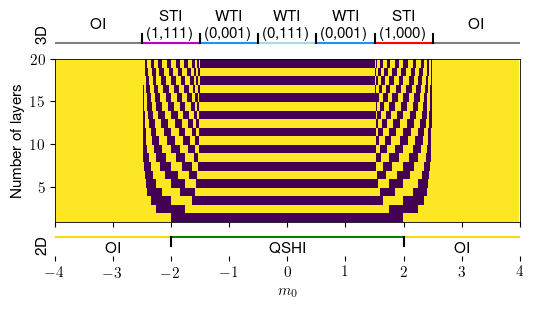}}
& \raisebox{-.5\height}{\includegraphics[width=8cm]{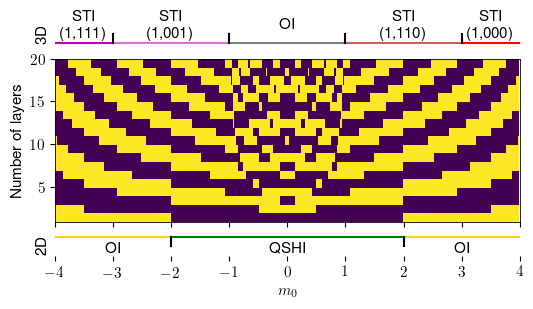}} & \rotatebox{90}{$b_z=3$}
\\
\rotatebox{90}{$b_z=1$} & \raisebox{-.5\height}{\includegraphics[width=8cm]{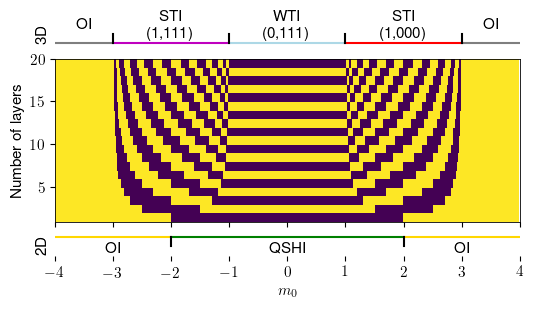}} & \raisebox{-.5\height}{\includegraphics[width=8cm]{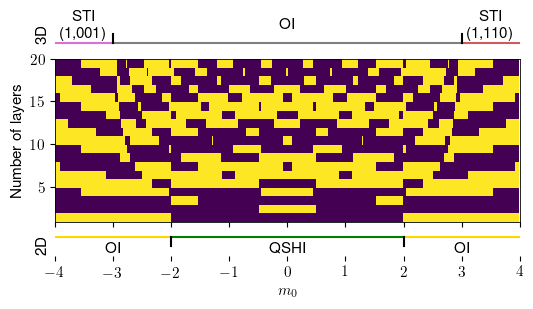}} & \rotatebox{90}{$b_z=5$}
\end{tabular}
\caption{Topological phase diagrams of the Wilson model for a slab as a function of its thickness and $m_0$ for $b_z=5, 3, 2, 1, 0.5, 0.001$ and $t_z = 0$. Yellow indicates a trivial ($\nu=0$) and dark blue a non-trivial ($\nu=1$) 2D topological invariant. Above and below each phase diagram are the 3D and 2D topological phase diagrams, respectively, for the corresponding value of $b_z$. Note that the colours of the 3D phase diagrams match the ones in Fig.\ \ref{fig:Bi2Se3-Wilson-3D-phase-diagram-comparison}.}
\label{fig:Wilson-Z2-map}
\end{figure*}

We now define a model which links the 2D and 3D limits by varying the number of layers $N_z$. It is formed of the two blocks \cite{Yoshimura2016}:
\begin{equation}
H_0(\textbf{k}) = \left( m_0 - b_\parallel \sum_{\mu = x, y} \cos(k_\mu) \right)\beta + \sum_{\mu = x, y} t_\mu \sin(k_\mu) \alpha_\mu
\label{eq:Wilson-H0}
\end{equation}
and
\begin{equation}
H_1(\textbf{k}) = -\frac{b_z}{2}\beta + i \frac{t_z}{2} \alpha_z,
\label{eq:Wilson-H1}
\end{equation}
which are combined into:
\begin{equation}
H =
\begin{pmatrix}
H_0 & H_1^\dagger & & & \\
H_1 & H_0 & H_1^\dagger & & \\
& H_1 & H_0 & & \\
& & & \ddots & H_1^\dagger\\
& & & H_1 & H_0
\end{pmatrix}.
\label{eq:Wilson-full-H}
\end{equation}
In the following, we choose $b_\parallel$ as the energy unit.

\section{\texorpdfstring{$\mathbb{Z}_2$}{Z2}-invariant maps}
\label{Z2-invariant maps}

The slab is infinitely extended in the planar $x$- and $y$-directions and finite in the $z$-direction. We can therefore picture the model in two different ways: either as a stack of two-dimensional four-band systems or as a single system with many bands. In the latter, we have a single 2D topological invariant $\nu$ for a given number of layers, which we calculate as \cite{Fu2007a}:
\begin{equation}
(-1)^\nu = \prod_{i=1}^4 \delta_i
\label{eq:nu}
\end{equation}
where the $\delta_i$ are products of the parity eigenvalues $\xi$ of the $N$ occupied energy bands at the four TRIM points $\Gamma_i$, namely $\Gamma$, $M$ and the two instances of $X$ on the $x$ and $y$ axes, which we name $X_x$ and $X_y$ respectively:
\begin{equation}
\delta_i = \prod_{m=1}^N \xi_{2m} (\Gamma_i)
\label{eq:delta-i}
\end{equation}
Note that the product is only over one band out of two. If it were not the case, since the bands are Kramers degenerate, we would always obtain $\delta_i=1$.

The three parameters entering the calculation of $\nu$ are $m_0$, $b_z$, and $t_z$. The first two also enter the calculation of the 3D topological invariant, and the third is an off-diagonal hopping amplitude in orbital space. We start in this section by studying the phase diagram for $t_z=0$. The influence of $t_z$ will be discussed in details in Sec.\ \ref{Impact of tz}.

Previously, three different regions were identified in the topological phase diagram of the Wilson-Dirac model in a slab geometry as a function of $m_0$ and number of layers, for fixed $b_z$ and $t_z$ \cite{Yoshimura2016}: an ordinary insulator (OI) region, a `stripe' region where $\nu$ alternates each time a layer is added, and a `mosaic' region between the two where $\nu$ oscillates with thickness. The two latter behaviours correspond to the non-oscillating and oscillating exponential decays of the surface gap identified in \cite{Ebihara2012}.

Here we calculate the topological, thickness dependent phase diagram for values of $b_z$ spanning the whole 3D phase diagram (Fig.\ \ref{fig:Wilson-Z2-map}). The phase diagram displays three types of regions: the OI regions where $\nu=0$ at any thickness for large $|m_0|$, the `stripe' region for small $|m_0|$ where $\nu=1$ for an odd and $\nu=0$ for an even number of layers, and the two `mosaic' regions around $|m_0|=2$, separating the OI and `stripe' regions. The latter are more complex and are characterised by zero and non-zero values of $\nu$ for any fixed thickness or $m_0$. In the limit where $b_z$ goes to zero, the `mosaic' regions shrink up to the point where they cannot be seen on the figure, although they have a finite size for any $b_z>0$. At $b_z=2$, the `mosaic' regions have expanded such that they merge at $m_0=0$. Upon increasing $b_z$ further, a new and even more complex region arises which supports, like the `mosaic' phase, non-zero values of $\nu$ at any thickness.

The 2D limit is naturally included in the slab model with the result for a single layer, i.e.\ thickness 1. In Fig.\ \ref{fig:Wilson-Z2-map}, we can witness the connection between the qualitatively different regions and the 3D limit discussed in \cite{Kobayashi2015}: the OI regions extrapolate to the 3D OI phase, the `stripe' regions to the weak topological insulator (WTI) phases, and the `mosaic' regions to the strong topological insulator (STI) phases. However, this extrapolation only holds for $t_z=0$ \cite{Kobayashi2015}. In addition, we find that the region at high $b_z$, which had not been discussed before, extrapolates to the 3D OI phase.

The phase diagram at low $b_z$ corresponds to the paradigmatic picture of stacking described for example in \cite{Fu2007a}: the stacking of 2D trivial insulators builds up a 3D trivial insulator, and stacking 2D quantum spin Hall insulators alternatively gives trivial and non-trivial systems for even and odd thicknesses respectively, and leads to a weak topological insulator in 3D. But as soon as we turn on $b_z$, the `mosaic' region arises, and either stacking OIs or QSHIs can extrapolate to an STI without apparent distinction. Moreover, for systems close enough to the boundary between STI and WTI or STI and OI, the system will reach the `mosaic' region only at large thicknesses, and therefore behaves like a paradigmatic stacking of QSHIs or OIs until it reaches this critical thickness. Strikingly different is the situation for the new region at large $b_z$. Indeed, there, the system features some topologically non-trivial states at any thickness but extrapolates towards a trivial insulator. At $b_z=5$ and $2<|m_0|<3$, there is even a region where stacking 2D trivial insulators extrapolates towards a 3D trivial insulator, but with $\nu$ irregularly switching between 0 and 1. The structure of the `mosaic' regions implies that thickness is a key parameter, particularly close to 3D topological phase transitions. Indeed, there one can witness boundaries between two regions as a function of thickness even at large thicknesses, meaning thicknesses that could be experimentally easy to access.

Here we have limited ourselves to visually identifying regions that display a similar behaviour with respect to $\nu$ versus $m_0$ and thickness. In the following, we explore the physical origin of the evolution of $\nu$ by decomposing $\nu$ into parity eigenvalues.

\section{Decomposition of the \texorpdfstring{$\mathbb{Z}_2$}{Z2}-invariant in parity eigenvalues}
\label{Decomposition of Z2-invariant in parity eigenvalues}

\begin{figure}
\centering
\includegraphics[width=8cm]{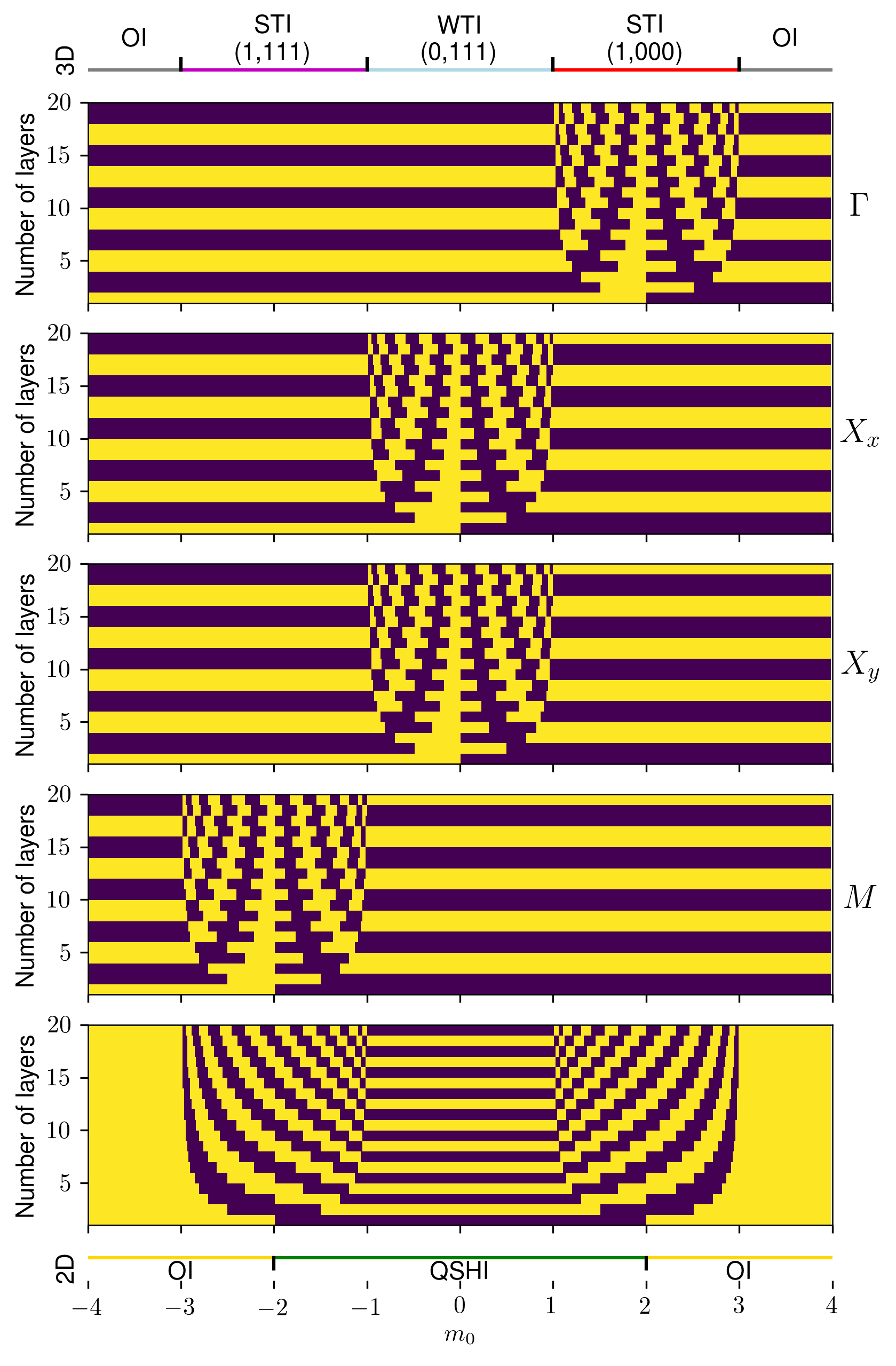}
\caption{Parity eigenvalues in the Wilson model for a slab at four time-reversal invariant points in the Brillouin zone: $\Gamma$, $X_x$, $X_y$ and $M$ (top four plots), and 2D topological invariant $\nu$ (bottom plot), as a function of $m_0$ and thickness for $b_z=1$ and $t_z =0$.}
\label{fig:Parity-eigenvalues-map}
\end{figure}

We return to the recipe of the calculation of $\nu$ via the parity eigenvalues at the four TRIM points, $\Gamma$, $X_x$, $X_y$ and $M$, using Eqs.\ \eqref{eq:nu} and \eqref{eq:delta-i} and evaluating the corresponding $\delta_i$ (Fig.\ \ref{fig:Parity-eigenvalues-map}).

For each TRIM point, we observe a single `mosaic' region centred on a specific $m_0$. For $\Gamma$, it is centred on $m_0=2$, for $X_x$ and $X_y$ it is centred on $m_0=0$, and for $M$ it is centred on $m_0=-2$. The product of the four $\delta_i \in \{-1,1\}$ gives $\nu$ following Eq.\ \eqref{eq:nu}. Since $X_x$, $X_y$ are twofold degenerate, their contribution is squared in Eq.\ \eqref{eq:nu}, and $\nu$ is determined by the $\Gamma$ and $M$ points only. Hence there is no `mosaic' region centred on $m_0=0$ in $\nu$. The two `mosaic' regions in $\nu$ centred on $m_0=2$ and $m_0=-2$ are associated with the TRIM points $\Gamma$ and $M$, respectively. In-plane anisotropies would lift the degeneracy of the $X$ points, and other topological phases would arise \cite{Yoshimura2014}.

\begin{figure}
\centering
\includegraphics[width=8.7cm]{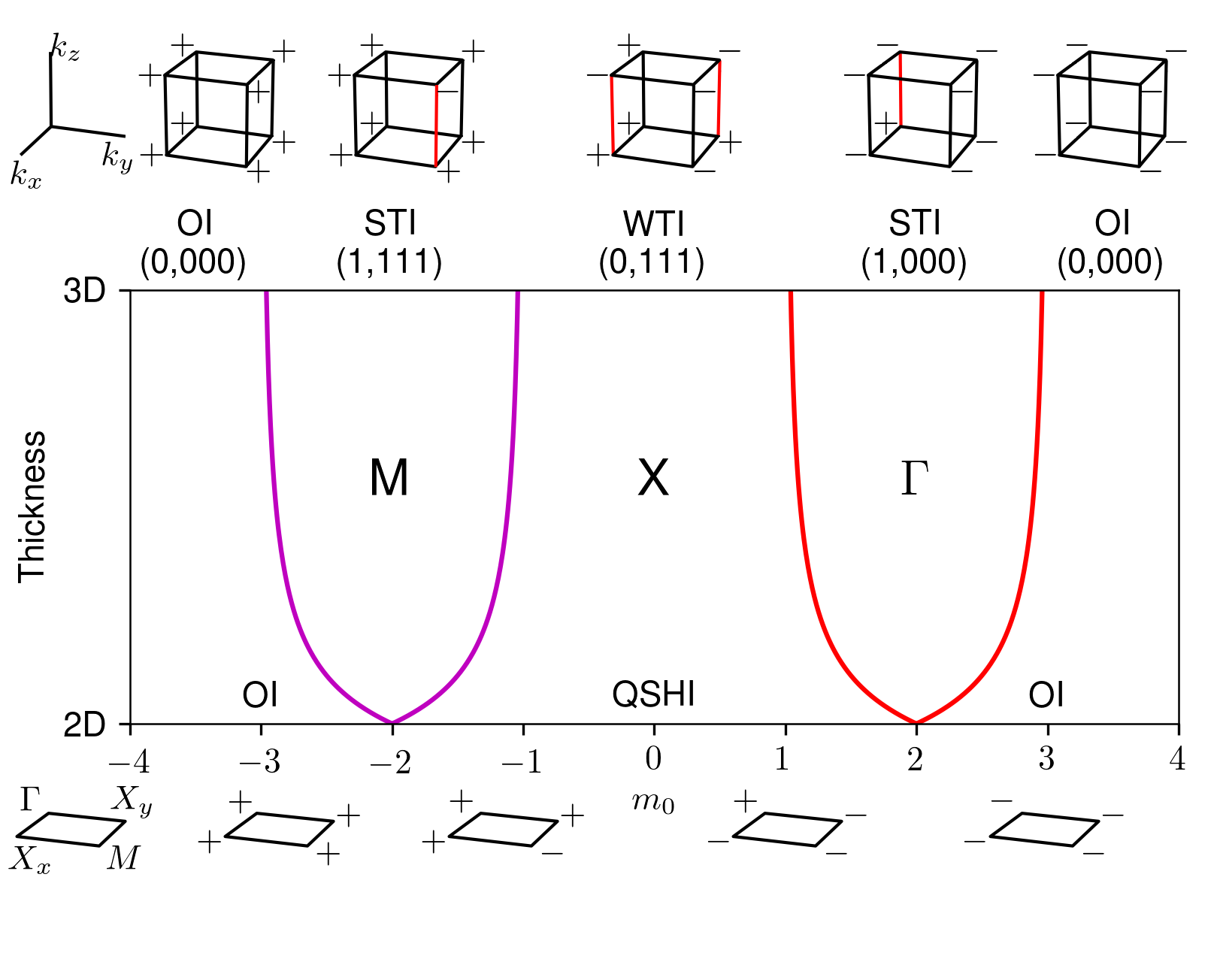}
\caption{Schematics of the evolution of the Wilson-Dirac model with thickness as a function of $m_0$ at $b_z=1$, with the parity cubes and parity squares in the 3D and 2D limits respectively. The red and purple lines denote the boundaries of the `mosaic' regions in the $\mathbb{Z}_2$-invariant map related to $\Gamma$ and $M$ respectively.}
\label{fig:Parity-eingenvalues-schematics}
\end{figure}

Since the slab is infinite in two dimensions and finite along $z$, we can place the four $\delta_i$ on a parity square consisting of the TRIM points in the Brillouin zone, just like in 2D. Besides the $\mathbb{Z}_2$-invariant maps we thereby have this additional tool to connect the parity cubes in 3D to the parity squares in 2D. In Fig.\ \ref{fig:Parity-eingenvalues-schematics}, we plot schematically the boundaries between the above encountered regions at $b_z=1$, along with the parity squares in 2D and parity cubes in 3D. We observe that the boundaries correspond to differences between top and bottom squares in the parity cubes. For example, when starting from a 2D system at $m_0 \in [2,3]$, the parity square bears only $-1$ entries. For the same values of $m_0$ in 3D, the parity cube has only $-1$ entries except for the $\Gamma$ point with parity $+1$. I.e., the bottom and the top square in this parity cube are different, which is highlighted by a red line in the corresponding parity cube in Fig.\ \ref{fig:Parity-eingenvalues-schematics}. This difference is related to the crossing of the boundary between the 2D and 3D limits. This boundary between $m_0=2$ and $m_0=3$ is in fact associated with $\Gamma$, as the TRIM point at which the difference between the two squares arises. Similar observations carry over to the whole range of $m_0$: at the TRIM point where top and bottom squares in the 3D parity cube differ the corresponding $\delta_i$ oscillate with thickness in a `mosaic' pattern.

This can be interpreted physically: if a system evolves from 2D to 3D, it acquires a third dimension in $\textbf{k}$-space. It starts from a system with a 2D Brillouin zone and ends with a 3D Brillouin zone. If the squares on top and bottom of the parity cube have different signs, the system cannot show this in the slab geometry since it has a 2D Brillouin zone. However in that case the corresponding $\delta_i$ oscillates with a peculiar pattern, indicating the conflict between top and bottom squares. The crossing of region boundaries at finite thickness therefore allows us to predict differences between the two parity squares in the 3D limit, and thus the 3D topological invariants.

The topological phase transitions for a fixed number of layers $N_z$ were previously obtained for the slab Hamiltonian in Eq.\ \eqref{eq:Wilson-full-H} \cite{Kobayashi2015}. The criterion for gap closing at a TRIM point leads to:
\begin{equation}
m_{2D}(\Gamma_i) = 2 \sqrt{\frac{b_z^2}{4} - \frac{t_z^2}{4}} \cos \left( k_z^{(n)} \right)
\label{eq:top-transitions}
\end{equation}
where $m_{2D}(\Gamma_i)$ is the coefficient in front of $\beta$ in Eq.\ \eqref{eq:Wilson-H0} evaluated at $\Gamma_i$, and
\begin{equation}
k_z^{(n)} = \frac{n \pi}{N_z + 1}, \quad n \in \llbracket 1, N_z \rrbracket
\label{eq-kz}
\end{equation}
where $\llbracket \rrbracket$ denotes a range of integers. The $k_z^{(n)}$ are obtained by diagonalising the Hamiltonian, similarly to an open chain problem. As such, they can be interpreted as `discrete $\textbf{k}$-point' along $z$. At the $\Gamma$ point:
\begin{equation}
m_{2D}(\Gamma) = m_0 - 2
\end{equation}
which translates into the values:
\begin{equation}
m_0 = 2 + \sqrt{b_z^2 - t_z^2} \cos \left( \frac{n \pi}{N_z + 1} \right)
\end{equation}
corresponding to topological phase transitions. If we let aside for the moment that $N_z$ is an integer, we can revert this expression and obtain:
\begin{equation}
N_z = \frac{n \pi}{\arccos \left( \frac{m_0 - 2}{\sqrt{b_z^2 - t_z^2}} \right)} - 1
\label{eq:Nz-as-a-function-of-m0-n=1}
\end{equation}
Eq.\ \eqref{eq:Nz-as-a-function-of-m0-n=1} enables to identify the boundaries of the `mosaic' region centred on $m_0=2$ with two specific values of $n$: $1$ and $N_z$. Via Eq.\ \eqref{eq-kz} we find:
\begin{align}
k_z^{(1)} = \frac{\pi}{N_z +1} \xrightarrow[N_z \to \infty]{} 0
\label{eq:n=1}
\\
k_z^{(N_z)} = \frac{N_z \pi}{N_z +1} \xrightarrow[N_z \to \infty]{} \pi
\label{eq:n=Nz}
\end{align}
The limits of $k_z^{(1)}$ and $k_z^{(N_z)}$ are particularly relevant, since they match the values that $k_z$ takes at TRIM points for infinite thickness, namely $0$ and $\pi$. Moreover, $k_z^{(1)}$ and $k_z^{(N_z)}$ each correspond to a single boundary: $k_z^{(1)}$ to the boundary at $m_0>2$ and $k_z^{(N_z)}$ to the boundary at $m_0<2$ and these boundaries are therefore respectively associated with $k_z=0$ and $k_z=\pi$ in the 3D limit. The same was calculated for $M$ and the `mosaic' region centred on $m_0=-2$. The boundary at $m_0>-2$ is associated with $k_z=0$ while the boundary at $m_0<-2$ is associated with $k_z=\pi$

If we come back to the earlier example of $m_0 \in [2,3]$, we see that the difference between the top and bottom squares of the 3D parity cube arises from the $\Gamma$ point, where there is one minus which differs from the 2D parity square. This corresponds precisely to the value $k_z=0$ to which we associated the region boundary at $m_0 \in [2,3]$. We therefore conclude: if there is full knowledge of the parity square in 2D, we can deduce from the region boundaries at finite thickness the exact 3D parity cube in the 3D limit, and therefore the topological invariants.

\section{Impact of \texorpdfstring{$t_z$}{tz}}
\label{Impact of tz}

\begin{figure}
\centering
\begin{tabular}{cc}
\rotatebox{90}{$t_z=0.7$} & \raisebox{-.5\height}{\includegraphics[width=8cm]{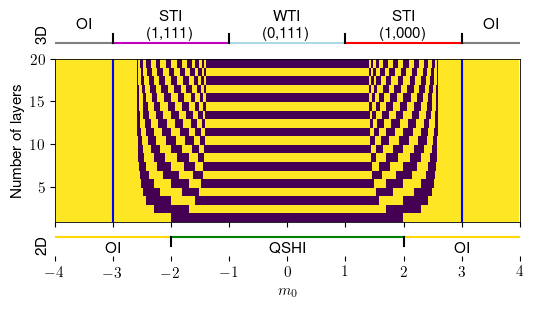}}
\\
\rotatebox{90}{$t_z=1.1$} & \raisebox{-.5\height}{\includegraphics[width=8cm]{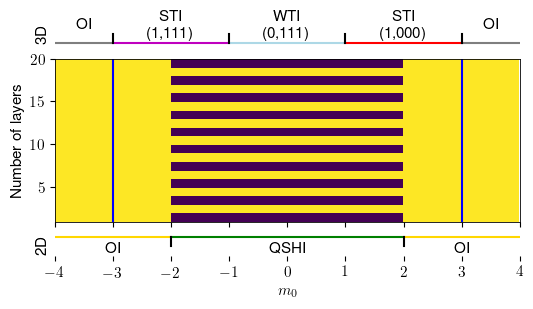}}
\end{tabular}
\caption{$\mathbb{Z}_2$ map at $b_z=1$ for $t_z=0.7$ and $t_z=1.1$. The corresponding map for $t_z=0$ is plotted in Fig.\ \ref{fig:Wilson-Z2-map}. The vertical blue lines delineate the range of $m_0$ for which there are gapless surface states in the limit of infinite thickness.}
\label{fig:Z2-map-tz}
\end{figure}

So far, we limited ourselves to the case $t_z=0$, in which the region boundaries extrapolate to the 3D topological phase transitions. However, it was calculated previously \cite{Kobayashi2015} that, as soon as we turn $t_z$ on, the `mosaic' regions in the 2D topological invariant maps do not extrapolate to the 3D topological transitions. This can directly be seen in Eqs.\ \eqref{eq:top-transitions}: when $t_z>0$, the coefficient in front of the cosine becomes smaller, up to the point where it vanishes for $b_z=t_z$. This is most striking: $t_z$ does not enter the calculation of the 3D topological invariants, but it plays a key role in the properties of the finite thickness system.

It was shown in \cite{Kobayashi2015} for $N_z=3,4$, that raising $t_z$ leads to the shrinking of the `mosaic' region and its disappearance at $t_z=b_z$. In order to scrutinise this further, we plot the evolution of the $\mathbb{Z}_2$ maps at $b_z=1$ for two values of $t_z$ (Fig.\ \ref{fig:Z2-map-tz}). As yielded by analytics, we see that the `mosaic' region shrinks when $t_z$ is increased, up to the point where $t_z=b_z$, where the `mosaic' region is completely suppressed, and only the OI and `stripe' regions are left. In this latter case, the $\mathbb{Z}_2$ map is the same as in the limit $b_z \to 0$ (Fig.\ \ref{fig:Wilson-Z2-map}).

\begin{figure}
\centering
\includegraphics[width=8cm]{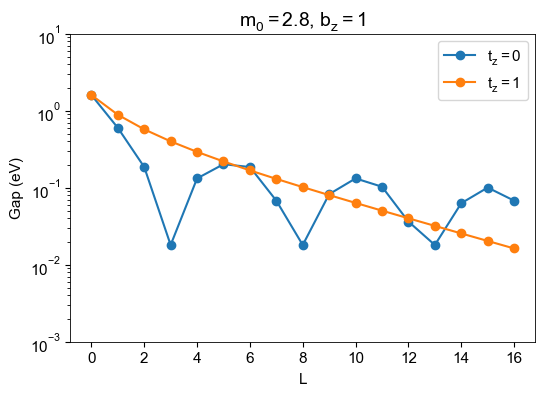}
\caption{Evolution of the surface gap with thickness at $m_0=2.8$ and $b_z=1$ for two different values of $t_z$.}
\label{fig:gap-tz}
\end{figure}

This motivates the question: since the `mosaic' and `stripe' regions have been associated respectively with oscillating and non-oscillating exponential decay, what is the consequence of the vanishing of the `mosaic' phase on the evolution of the surface gap? This was studied in a one-dimensional model at the $\Gamma$ point, which found that the gap evolution with thickness changes radically when changing $t_z$, going from oscillating behaviour to non-oscillating behaviour \cite{Okamoto2014}. We confirmed this in the full Wilson-Dirac model by plotting the evolution of the surface gap with thickness at $m_0=2.8$ and $b_z=1$, for two different values of $t_z$ (Fig.\ \ref{fig:gap-tz}). At these values of $m_0$ and $b_z$, the system is inside the `mosaic' region for $t_z=0$ and outside of it for $t_z=1$. We confirm that the decay is oscillating in the first case and non-oscillating in the second case. The association of the oscillating behaviour with the `mosaic' phase is therefore consistent with what we calculated.

Very importantly, the non-oscillating behaviour obtained at high $t_z$ still decreases exponentially with thickness, and converges to zero in the limit of infinite thickness. This fits the results in \cite{Okamoto2014}, which showed that the surface gap goes to zero independently of $t_z$. When the 3D topological invariant is such that we expect surface states on the top and bottom surfaces (i.e.\ in the case of a strong topological insulator, or in the case of a weak topological insulator with $\nu_1\nu_2\nu_3 \neq 001$), then the surface gap goes to zero when $L$ goes to infinity, for any value of $t_z$. Therefore, even though the $\mathbb{Z}_2$ map is entirely changed by $t_z$, the existence of surface states is independent of it, and still entirely consistent with the 3D topological numbers. In particular, there is a region in the $\mathbb{Z}_2$ map for $t_z>0$, between the boundary of the `mosaic' region and the value of $m_0$ corresponding to the 3D topological phase transition, where there are simultaneously surface states and a constant value $\nu=0$.

Therefore, $t_z$ is of key importance for the evolution of the Wilson-Dirac model with thickness. It is, in the 3D model, (Eq.\ \eqref{eq:3D-Wilson-model}), the perpendicular Dirac velocity. And it becomes, in the slab model (Eqs.\ \eqref{eq:Wilson-H0}, \eqref{eq:Wilson-H1} and \eqref{eq:Wilson-full-H}), the strength of the pure imaginary hopping along $z$. Unlike $b_z$, which  plays a role in the calculation of topological invariants in the 3D model as well as in the slab model, $t_z$ only affects the topological invariant in the slab model. A different way to say this is that at $t_z=0$ the change of boundary conditions does not affect the topological invariant, while when $t_z>0$ it does affect it.

This brings us back to the issue about the bulk-boundary correspondence presented in the introduction: can one relate the existence of gapless states at the boundary between two half-spaces to their intrinsic properties, and not to the topological invariant of the corresponding 3D space? First, we notice that our geometry is different here, since we do not have a half-space but a slab, with two surfaces and not one. And in this geometry, we can answer negatively: we are unable to find an intrinsic property of our slab predicting the occurrence of surface states.

In this section, we have shown that there are two different types of evolution with thickness from a 2D system to a strong topological insulator. One at low $t_z$, where the surface gap oscillates and where the first drop in this gap corresponds to the crossing of the phase transition we defined in the previous section, and one where the surface gap decreases exponentially and never crosses that transition. The natural question to ask now is: how to differentiate physically between these two evolutions?

\section{Band inversion}
\label{Band inversion}

\begin{figure}
\centering
\includegraphics[width=8cm]{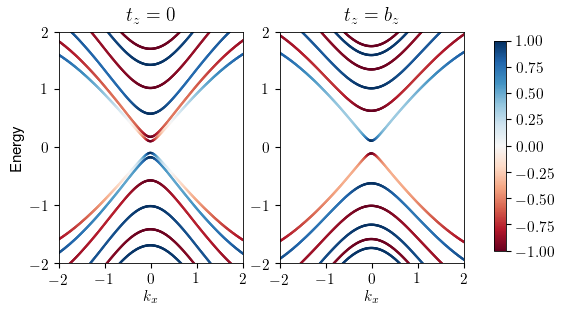}
\caption{Band structure at $m_0=2.8$ and $b_z=1$ for two different values of $t_z$. In the first case, there is band inversion, while there is not band inversion in the second case.}
\label{fig:Band-inversion}
\end{figure}

In order to differentiate between the two types of evolution from a 2D system to a strong topological insulator discussed in the previous section, we come back to the physical meaning of the region boundaries we calculated in Sec.\ \ref{Decomposition of Z2-invariant in parity eigenvalues}. Every change in $\nu$ corresponds to a topological transition at one of the $k_z^{(n)}$, i.e.\ the discrete $\textbf{k}$-points along $z$. More specifically, they correspond to a change in the parity eigenvalues at $(k_x,k_y)=\Gamma,M$. In order for such a change to happen, it is required that the bands invert at the transition. Therefore the transition calculated in Sec.\ \ref{Decomposition of Z2-invariant in parity eigenvalues} corresponds to band inversion. Inside the region which it delimits the bands are inverted, and outside they are not.

We can visualise this by defining the parity without $\textbf{k}$-flip:
\begin{equation}
PWkF(n,\textbf{k}) = \sum_{i = 1}^{N_z} \psi_{N_z - i} (n,\textbf{k})^\dagger \beta \psi_i(n,\textbf{k})
\end{equation}
where $\psi$ is an eigenvector, $n$ is a band index and $i$ is a layer index. It reduces to the parity at the four 2D TRIM points, and thus allows to visualise the change of parity from one TRIM point to another.

We plot its evolution in the band structure in Fig.\ \ref{fig:Band-inversion}. We find that at $t_z=0$ the top valence band changes sign of PWkF from small $k_x$ to large $k_x$. This indicates that for this band, parity is $-1$ on the edge of the Brillouin zone and $+1$ at $\Gamma$. The same observation holds with the opposite sign for the bottom conduction band. This can be understood by an exchange of parity character between the two bands closest to the gap, i.e.\ by saying that there is band inversion close to the $\Gamma$ point. This fits the $\mathbb{Z}_2$-map in Sec.\ \ref{Decomposition of Z2-invariant in parity eigenvalues}: for this set of parameters, we are inside the `mosaic' region. At $t_z=1$, the PWkF remains constant along every single band, and therefore no band inversion is observed. This fits the $\mathbb{Z}_2$-map for this value of $t_z$ (Fig.\ \ref{fig:Z2-map-tz}) at which there is no more `mosaic' region, and in which, for this set of parameters, we are inside the OI region. At both values of $t_z$, the surface gap is small, as expected from its evolution calculated in the previous section.

Therefore the region boundary we calculated in Sec.\ \ref{Decomposition of Z2-invariant in parity eigenvalues} is a topological phase transition between the band inverted and non-band inverted regimes. In particular, for $t_z>0$, there is a region of the phase diagram which converges towards a 3D topological insulating state with gapless surface states, but does not have band inversion at any finite thickness, and is topologically trivial in terms of the 2D topological invariant.

\begin{figure}
\centering
\includegraphics[width=8cm]{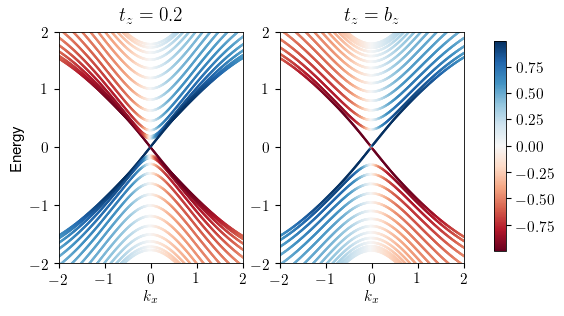}
\caption{Band structure at $m_0=2.8$ and $b_z=1$ for two different values of $t_z$. In both cases, the surface states are momentum-locked.}
\label{fig:Spin-projection}
\end{figure}

This yields the question: are these surface states topological? Indeed, some other systems are known to host surface states without connection to topology. One example of this is diamond \cite{Graupner1997}. One can look for particular features of topological surface states, such as their helical character, i.e.\ the fact that they exhibit momentum-spin locking \cite{Qi2011}. In order to do this, we define the spin texture following \cite{Salehi2016}:
\begin{equation}
ST(n,\textbf{k}) = \sum_{i = 1}^{N_z} \psi_{i} (n,\textbf{k})^\dagger \alpha_x \psi_i(n,\textbf{k})
\end{equation}
Its evolution in the band structure is plotted in Fig.\ \ref{fig:Spin-projection}, for two different values of $t_z$. In both cases, the surface states are indeed each related only to one spin, as expected for topological surface states. This is however not sufficient to conclude without ambiguity that these states originate from the topology of the system, since spin-momentum locked states have been observed on the surface of non-topological materials such as gold \cite{LaShell1996}. Nevertheless, these surface states exist only when the parameters of the slab system correspond to a non-trivial 3D phase. This in itself is a strong indication of their topological origin.

Coming back to the issue of predicting 3D topological invariants from finite-thickness systems, we can now conclude that at $t_z=0$, one can predict directly the topological invariant in 3D from the 2D parity square and the transitions observed at finite thickness, while for $t_z>0$ one can only draw conclusions from crossing a transition, and not from the fact that no transition was crossed at any finite thickness.

\section{Influence of \texorpdfstring{$t_z$}{tz} in the 3D limit}
\label{Influence of tz in the 3D limit}

In the previous two sections, we have established that $t_z$ is of key importance since it tunes the position of the topological phase transition between the inverted and non-inverted regimes. From the definition of the 3D Wilson-Dirac Hamiltonian (Eq.\ \eqref{eq:3D-Wilson-model}), we know that $t_z$ is the strength of the Dirac term along $z$, i.e. it is the linear component of hopping along $z$ close to $\Gamma$. Because of this, its term is zero at the TRIM points and it does not enter the calculation for the 3D topological invariants.

\begin{figure}
\centering
\includegraphics[width=8cm]{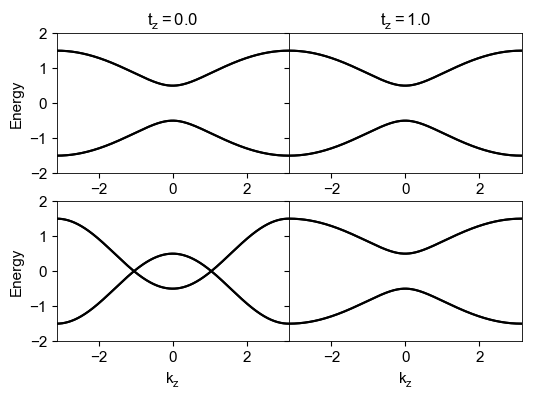}
\caption{Band structure of the 3D Wilson model at $m_0=2.5$ and $b_z=1$ for two different values of $t_z$, with $k_y=k_z=0$ (top) and $k_x=k_y=0$ (bottom). Note that $t_z$ opens a gap along $k_z$.}
\label{fig:3D-band-structure-tz}
\end{figure}

However, in the 3D model, changing $t_z$ influences the band structure. The evolution of the band structure as a function of $t_z$ is particularly important to understand the crossover from 2D to 3D of our system at an arbitrary value of $t_z$. We calculated the band structure at $m_0=2.5$ and $b_z=1$ for two values of $t_z$ (Fig.\ \ref{fig:3D-band-structure-tz}). We find that the variation of $t_z$ has no influence on the band structure along $k_x,k_y$, but has a strong influence on the band structure along $k_z$. Indeed, at $t_z=0$, the gap closes in two points along $k_z$, and the system is not an insulator. Hence, in the only limit where the topological phase transitions at finite thickness extrapolate to the 3D topological phase transitions, the system is not a topological insulator in 3D any more. Increasing $t_z$ opens the gap, and the spectrum along $k_z$ becomes the same as along $k_x$ and $k_y$ at $t_z=1$.

\begin{figure}
\centering
\includegraphics[width=8cm]{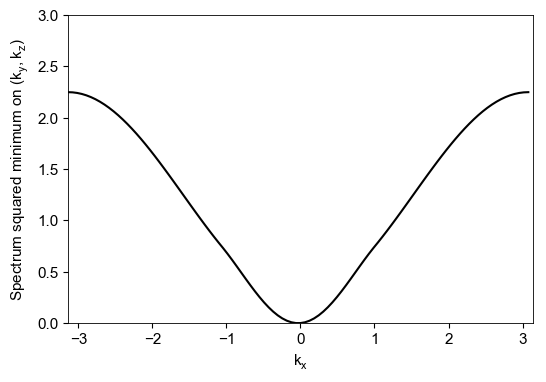}
\caption{Minimisation of the band structure for all $k_y$, $k_z$ as a function of $k_x$ at $m_0=2.5$, $b_z=1$ and $t_z=0$. The band structure crosses the Fermi level only when $k_x=0$.}
\label{fig:3D-band-structure-minimum}
\end{figure}

In order to verify whether there are other gap closing points, we calculated, for each point along the axis $k_y=k_z=0$, the minimum of the square of the band structure along the plane perpendicular to this axis and going through this particular point (Fig.\ \ref{fig:3D-band-structure-minimum}). This shows that, for any value of $k_y$ and $k_z$, the gap only ever reaches zero at $k_x=0$. Since the $k_x$ and $k_y$ axes are symmetric, we also conclude that the gap only closes at $k_y=0$. Therefore, the two gap-closing points shown on Fig.\ \ref{fig:3D-band-structure-tz} are the only two points where the gap closes in the entire Brillouin zone.

One important consequence of the fact that the system is a metal at $t_z=0$ is that if a system is an insulator in the bulk, then $t_z$ must be larger than zero, and therefore the new region discussed in Sec.\ \ref{Band inversion}, which displays surface states but no band-inversion for the slab, is present in some range of $m_0$. Hence, tuning the Dirac mass $m_0$ of a topological insulator leads to finding this new region.

This observation carries on to large portions of the phase diagram as a function of $m_0$ and $b_z$. This can be calculated by setting $t_z=0$ in Eq.\ \eqref{eq:3D-Wilson-model} and looking for gap-closing points along $k_x=k_y=0$. We find gap-closing points everywhere except in the OI phases at large $|m_0|$ and in the weak topological phases with weak indices $\nu_1\nu_2\nu_3 = 001$.

In the previous sections, we have identified the region boundaries with topological phase transitions separating inverted and non-inverted band structures. We now study in details the extreme case encountered at high $b_z$ in Sec.\ \ref{Z2-invariant maps} and study whether the picture developed in the previous sections fits this peculiar case.

\section{Ordinary insulating phase at high \texorpdfstring{$b_z$}{bz}}
\label{Ordinary insulating phase at high bz}

\begin{figure}
\centering
\includegraphics[width=8cm]{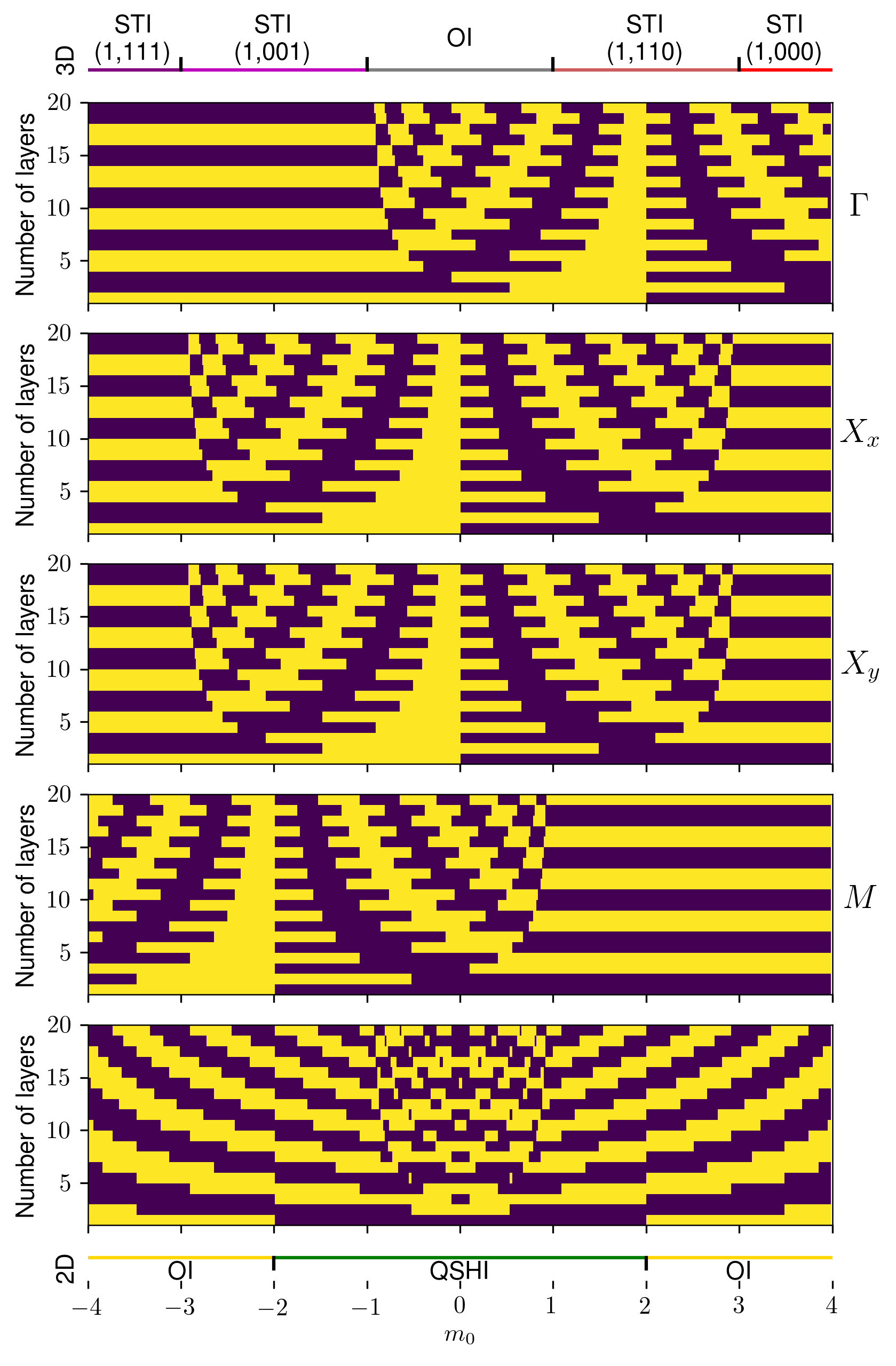}
\caption{Parity eigenvalues in the Wilson model for a slab at four time-reversal invariant points in the Brillouin zone: $\Gamma$, $X_x$, $X_y$ and $M$ (top four plots), and 2D topological invariant $\nu$ (bottom plot), as a function of $m_0$ and thickness for $b_z=3$ and $t_z =0.5$.}
\label{fig:Parity-eigenvalues-map-high-bz}
\end{figure}

We now focus on the new region found in Sec.\ \ref{Z2-invariant maps} at high $b_z$, centred on $m_0=0$ with zero and non-zero $\nu$ at any thickness and extrapolating towards a 3D OI state. We follow the reasoning in Sec.\ \ref{Decomposition of Z2-invariant in parity eigenvalues} and calculate the four $\delta_i$ at high $b_z$ (Fig.\ \ref{fig:Parity-eigenvalues-map-high-bz}). We find that each $\delta_i$ behaves very much like at lower $b_z$: it oscillates with $m_0$ at finite thickness, or with thickness at fixed $m_0$, in one region centred on a particular value of $m_0$. However, the width of this region is much larger than at lower $b_z$. This fits our results in Fig.\ \ref{fig:Wilson-Z2-map}, where the `mosaic' regions is growing in size when $b_z$ is enlarged.

As before, the two $\delta_i$ associated with $X_x$ and $X_y$ do not play a role since they are equal and thus only multiply $\nu$ by one. The peculiar region close to $m_0=0$ arises from the product of oscillating regions at two TRIM points: $\Gamma$ and $M$. This in particular explains that such a region was not seen at lower values of $b_z$: it can only exist when the width of the two oscillating regions in $\delta_\Gamma$ and $\delta_M$ is larger than their separation.

We proceed to check the validity of our analysis in Sec.\ \ref{Decomposition of Z2-invariant in parity eigenvalues}. According to it, crossing a region boundary corresponds to having a difference between top and bottom squares of the parity cube in the 3D limit. In the range of $m_0$ in which the new region is present, we notice that we cross four different region boundaries when increasing thickness, one for each TRIM point. Hence, the prediction is that the top and bottom squares of the parity cubes are entirely different. This is indeed what is found in the 3D limit: the bottom square only has $+1$ entries, and the top square only $-1$ entries. Consequently all the 3D topological numbers are zero, i.e.\ the system extrapolates to a 3D ordinary insulator, fitting our calculation for the 3D phase diagram. Our analysis in Sec.\ \ref{Decomposition of Z2-invariant in parity eigenvalues} is therefore confirmed to be valid even in this case.

\begin{figure}
\centering
\includegraphics[width=8cm]{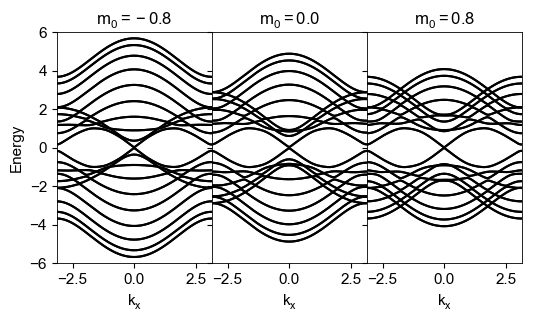}
\caption{Band structure at $b_z = 3$ and $t_z = 0.5$ for three different values of $m_0$.}
\label{fig:3D-band-structure-high-bz}
\end{figure}

Beyond this, the region at high $b_z$ has the most peculiar feature of having non-zero values of $\nu$ at any thickness for some values of $m_0$. Even at very high thicknesses, the system keeps on being non-trivial in some parameter range even though it extrapolates to an OI phase in 3D. The results at $b_z=5$ (Fig.\ \ref{fig:Wilson-Z2-map}) are even more striking since they feature, for $2<|m_0|<3$, a region where one can go from 2D OI to 3D OI by crossing this large region centred on $m_0=0$, which at this value of $b_z$ is even larger. This means that, in some range of $m_0$, stacking a 2D ordinary insulator gives an ordinary insulator in the 3D limit, but still features non-trival values of $\nu$ at finite thickness.

Since the `mosaic' regions are associated, in the limit $t_z \to 0$, with the presence of surface states, it is legitimate to ask whether the new region at high $b_z$ does not also feature surface states, since it also has $\nu$ oscillating with thickness. We calculated the band structure in this state at three different values of $m_0$ (Fig.\ \ref{fig:3D-band-structure-high-bz}). We indeed see bands close to the Fermi level both in the center and on the edge of the Brillouin zone. This is true for every value of $m_0$ calculated, the main difference in the band structure when varying $m_0$ being how much the bands away from the Fermi level are grouped together.

\begin{figure}
\centering
\includegraphics[width=8cm]{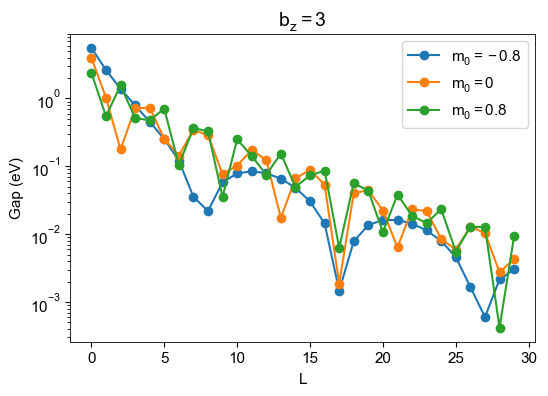}
\caption{Evolution of the surface gap with thickness at $b_z = 3$ and $t_z = 0.5$ for three different values of $m_0$.}
\label{fig:Gap-high-bz}
\end{figure}

These bands close to the Fermi level are however not entirely gapless, which is expected since we have a finite size system along $z$. We calculated the surface gap as a function of the number of layers for the same parameters as for the five spectra in the previous figure (Fig.\ \ref{fig:Gap-high-bz}). We find that the gap decays exponentially with thickness, as expected for surface states. Moreover, this decay is not strictly exponential, but also features an oscillating component. On Fig.\ \ref{fig:Gap-high-bz}, this is most clear for $m_0=-0.8$, while for $m_0 \geq 0$ the variation around the exponential decay becomes harder to describe. This oscillating behaviour is very close to the one encountered in the `mosaic' regions, as plotted for example in Fig. \ref{fig:gap-tz}.

This new region at high $b_z$ is therefore associated with surface states behaving similarly to the ones of a strong topological insulator, but localised on the surface of an ordinary insulator. Moreover, as can be seen in the $\mathbb{Z}_2$ map for $b_z=5$ in Fig.\ \ref{fig:Wilson-Z2-map}, it can be found at values of $m_0$ on both sides of the topological phase transitions of the 2D Wilson-Dirac model between OI and QSHI states. Therefore, this state can be obtained by stacking 2D ordinary insulators or 2D topological insulators alike.

This behaviour arises only when $b_z$ is large which means that it is associated with strong hopping along $z$. Indeed, the $b_z$ term in the 3D model (Eq.\ \eqref{eq:3D-Wilson-model}) is the strength of the dispersion of the mass term. Systems which could display this interesting behaviour should therefore be strongly one-dimensional, unlike most experimentally studied topological insulators which are made of van der Waals-coupled layers and therefore two-dimensional. This can be very clearly seen in the position of Bi$_2$Se$_3$ on the 3D phase diagram on Fig.\ \ref{fig:Bi2Se3-Wilson-3D-phase-diagram-comparison}: it has a very small value of $b_z$, which fits its two-dimensional character.

\section{Conclusion}

We have studied in detail the evolution of the Wilson-Dirac model in a slab geometry as a function of thickness. First, in Sec.\ \ref{Z2-invariant maps}, we calculated the 2D topological invariant of a slab $\nu$ in the whole 3D topological phase diagram as a function of thickness at $t_z=0$. This allowed us to see the evolution of three regions of these slab topological phase diagrams that had been seen before: ordinary insulator (OI), `stripe' and `mosaic', converging to the 3D OI, weak and strong topological insulator phases, respectively. It also yielded a new region not calculated before at high $b_z$, characterised by a complex oscillation between $\nu=0$ and $\nu=1$, and converging towards a 3D OI phase.

In Sec.\ \ref{Decomposition of Z2-invariant in parity eigenvalues}, we decomposed $\nu$ into its components associated to TRIM points $\delta_i$. This determined that the two `mosaic' regions were associated each to one fluctuating region, either in $\delta_\Gamma$ or in $\delta_M$. Positioning the four $\delta_i$ on a parity square allowed us to relate the three regions in the $\mathbb{Z}_2$-invariant maps at low $b_z$ to differences between parity cubes in 3D and parity squares in 2D. We showed that the outermost transition of each `mosaic' region was related to a change of the parity square of the slab at the TRIM point related to the transition, and moreover that the ones at higher (respectively lower) $m_0$ were related to a change in the parity cube compared with the parity square in the 2D limit at $k_z=0$ (respectively $k_z=\pi$). Therefore, the crossing of these region boundaries determines entirely the 3D limit, and it is possible to predict the full 3D topological state from the crossing of these topological phase transitions at finite thickness.

We then turned to the influence of $t_z$, which had been set to zero in the previous sections, and which is responsible for shrinking the size of the `mosaic' regions, thus creating a region of the phase diagram that converges towards a strong topological insulating state but has $\nu=0$ at any thickness. In Sec.\ \ref{Impact of tz}, we calculated the evolution of the surface gap with thickness in this region, and found that it changes from an oscillating to a non-oscillating exponential decay when $t_z$ is increased, confirming previous results in a 1D model. More importantly, in both cases, the gap extrapolates to zero, meaning that there are surface states even in the absence of a non-trivial value of $\nu$ in this region. This yielded the question: what is the difference between the `mosaic' region and this new region, which both converge to a 3D topological insulating state and feature surface states. In Sec.\ \ref{Band inversion}, we answered this question by finding that the topological phase transition identified in Sec.\ \ref{Decomposition of Z2-invariant in parity eigenvalues} is the transition between band inverted and non-band inverted spectra. Therefore there is a parameter regime without band inversion at any thickness that displays surface states and converges towards a strong topological insulating phase.

In Sec.\ \ref{Influence of tz in the 3D limit}, we studied the band structure of the 3D system as a function of $t_z$, and found that at $t_z=0$ the system is gapless in the bulk, and therefore not a topological insulator. Meaning in particular that a strong topological insulator must have finite $t_z$ and thus exhibit the non-inverted region for the slab discussed in Sec.\ \ref{Band inversion} at some value of $m_0$. Finally, in Sec.\ \ref{Ordinary insulating phase at high bz}, we studied the new region at high $b_z$ found in Sec.\ \ref{Z2-invariant maps} and confirmed that it is also naturally understood using the picture defined in Sec.\ \ref{Decomposition of Z2-invariant in parity eigenvalues}. We moreover calculated that it has surface states whose gap decreases exponentially with thickness, even though it extrapolates towards a 3D ordinary insulator.

This work is of particular relevance to materials close to a topological phase transition. Indeed, there, the crossing of region boundaries happens at large thickness. Moreover, that is where the new phase with surface states but without band inversion discussed in Sec.\ \ref{Band inversion} arises. This means in particular that one could observe a gap-closing transition with thickness for systems sufficiently close to a 3D topological phase transition. Such a thickness-tuned transition is reminiscent of recent works on ZrTe$_5$ \cite{Lu2017, Li2018}, especially since it was calculated to be close to the transition between a weak and a strong topological insulating phases \cite{Weng2014, Fan2017}.

\section*{Acknowledgements}

We thank Daniel Braak and Patrick Seiler for stimulating discussions. This work was supported by the Deutsche Forschungsgemeinschaft --- Grant  number 107745057, TRR 80.

\bibliographystyle{apsrev4-1}
\bibliography{FSTI}

\end{document}